\documentclass{elsart}

\usepackage{natbib}

\usepackage{graphicx}

\usepackage{amssymb}

\newcommand{\vecvar}[1]{\mbox{\boldmath$#1$}}

\begin{document}

\begin{frontmatter}



\title{First-principles study on field evaporation of surface atoms from W(011) and Mo(011) surfaces}


\author{Tomoya Ono}
\address{Research Center for Ultra-Precision Science and Technology, Osaka University, Suita, Osaka 565-0871, Japan}
\author{Takashi Sasaki, Jun Otsuka and Kikuji Hirose}
\address{Department of Precision Science and Technology, Osaka University, Suita, Osaka 565-0871, Japan}

\begin{abstract}
The simulations of field-evaporation processes for surface atoms on W(011) and Mo(011) surfaces are implemented using first-principles calculations based on the real-space finite-difference method. The threshold values of the external electric field for evaporation of the surface atoms, which are $\sim$ 6 V/\AA \hspace{2mm} for tungsten and $\sim$ 5 V/\AA \hspace{2mm} for molybdenum, are in agreement with the experimental results. Whereas field evaporation has been believed to occur as a result of significant local-field enhancement around the evaporating atoms, in this study, the enhancement is not observed around the atoms but above them and the strength of the local field is much smaller than that expected on the basis of the classical model.
\end{abstract}


\end{frontmatter}

\section{Introduction}
In the last decade, atomic-scale manipulation on material surfaces using scanning tunneling microscopy has been the subject of intensive research due to its potential for creating artificial surface nanostructures. There have been several experiments in which an atom is transferred between a probe tip and a surface under the influence of applied voltage pulses \cite{a}. This phenomena is called field evaporation \cite{tsong} and is considered to be a thermally activated process in which rate constants can be parameterized according to the Arrhenius formula \cite{Arrhenius}. Since the activation energy for field evaporation varies depending on the electric field applied between the probe tip and surface, the contribution of the electric field is important during this process. In the past, Tsong \cite{tsong} proposed phenomenological models for evaluating the activation energy for the field evaporation of surface atoms. However, for a profound understanding of this phenomenon from a microscopic point of view, it is mandatory to self-consistently calculate both the electronic charge distribution and the induced electrostatic field. So far, first-principles calculations has been used to interpret the filed evaporation of the surface atoms on semiconductor surfaces \cite{watanabe, ono}. When it comes to the metal surface, although several first-principles studies have been implemented \cite{lang1,lang2,wang}, all of them employed {\it structureless jellium surfaces}.

In this study, first-principles calculations were carried out to explore the activation energies and threshold values of external electric fields for the evaporation of surface atoms from {\it atomically flat W(011) and Mo(011) surfaces}. We determined activation energies for the evaporation of atoms on atomically flat (011) surfaces under strong electric fields. Our calculated threshold values for field evaporation, $\sim$ 6 V/\AA \hspace{2mm} for tungsten and $\sim$ 5 V/\AA \hspace{2mm} for molybdenum, are in good agreement with experimental data. Whereas field evaporation has been believed to occur as a result of significant local-field enhancement (LFE) around single metal atoms on atomically flat crystal planes, we observed the LFE not around the evaporating atoms but $\sim$ 3 a.u. above them.

The rest of the paper is organized as follows: In Sec.~\ref{sec:cd}, we briefly describe the computational details of our calculations. The results are presented and discussed in Sec.~\ref{sec:rd}. We conclude our findings in Sec.~\ref{sec:concl}.

\section{Computational details}
\label{sec:cd}
\subsection{Method}
Our first-principles simulations are based on the real-space finite-difference method \cite{chelikowsky1,chelikowsky2} with the incorporation of the timesaving double-grid technique \cite{tsdg}. Compared with the plane-wave approach, the real-space finite-difference method is much simpler to implement while maintaining a high degree of accuracy. Moreover, the real-space calculations eliminate the serious drawbacks of the conventional plane-wave approach, such as its inability to describe strictly nonperiodic systems: In the case of simulations under external electric fields using the conventional plane-wave approach, the periodic boundary condition gives rise to a saw-tooth potential, which sometimes leads to numerical instability during the self-consistent iteration, while the real-space finite-difference method is free from involving the saw-tooth potential and can exactly determine the potential induced by the external electric field as a boundary condition. The grid spacing is set to be $\sim$ 0.33 a.u. which corresponds to a cutoff energy of $\sim$ 89 Ry in the plane-wave approach and a denser grid spacing of $\sim$ 0.11 a.u. is taken in the vicinity of nuclei with the augmentation of double-grid points \cite{tsdg}. The nine-point finite-difference formula is used for the derivative arising from the kinetic-energy operator of the Kohn-Sham equation \cite{hks2}. The norm-conserving pseudopotentials \cite{pseudop,kobayashi} incorporating a Kleinman-Bylander nonlocal form \cite{kleinman-bylander} are employed to describe the electron-ion interaction. Exchange-correlation effects are treated with the local-spin-density approximation \cite{lda} of the density functional theory \cite{hks1} .

\subsection{Models}
\begin{figure}[htb]
\begin{center}
\includegraphics{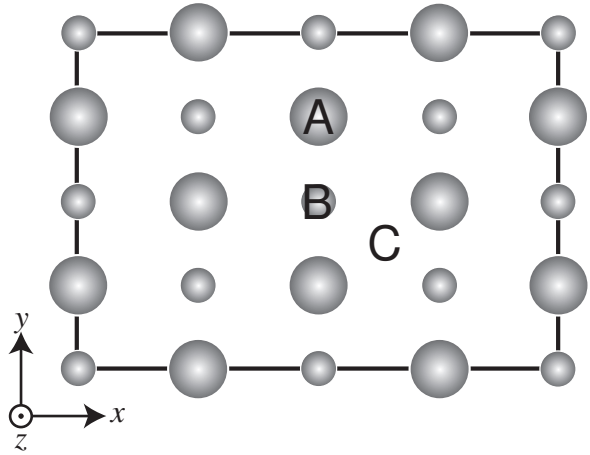}
\caption{Top view of the two topmost layers of atomically flat (011) surface. Large and small spheres represent atoms on the top and second layers, respectively.}
\label{fig:1}
\end{center}
\end{figure}

Figure \ref{fig:1} shows the top view of the (011) surfaces employed here. We adopt a technique that involves the use of a supercell whose size is set to be $L_x=2\sqrt{2}a_0$, $L_y=2a_0$ and $L_z=10a_0$, where $L_x$, $L_y$ and $L_z$ are the lengths of the supercell in the $x$, $y$ and $z$ directions, respectively. Here, the direction perpendicular to the surface was chosen as the $z$ direction and the bulk constants $a_0$ are 5.97 a.u. for tungsten and 5.43 a.u. for molybdenum. To completely eliminate unfavorable effects from neighboring cells which are artificially repeated in the case of the periodic boundary condition, the {\it nonperiodic} boundary condition of vanishing wave functions out of the supercell is imposed in the $z$ direction, while the periodic boundary conditions are employed in the $x$ and $y$ directions. The surface Brillouin zone is sampled by the uniform mesh of $\vecvar{k}$ points and the Fermi level is broadened by the Fermi distribution function. The supercell contains three (011) atomic layers (i.e., the thin film model) and the atoms in the topmost atomic layer are fully relaxed during structural optimizations.

We first determine the minimum-energy atomic configuration of a surface atom. A tungsten (molybdenum) surface atom is initially placed at each of the sites of A, B, and C in Fig. \ref{fig:1} at an appropriate distance from the W(011) [Mo(011)] surface, and then the forces on the atoms are relieved. The binding energies are listed in Table \ref{tbl:1}. The most stable site is found to be B on both the W(011) and the Mo(011) surfaces and hereafter we use this configuration as the computational model.

\section{Results and discussion}
\label{sec:rd}
\subsection{Activation energy}
\begin{figure}[htb]
\begin{center}
\includegraphics{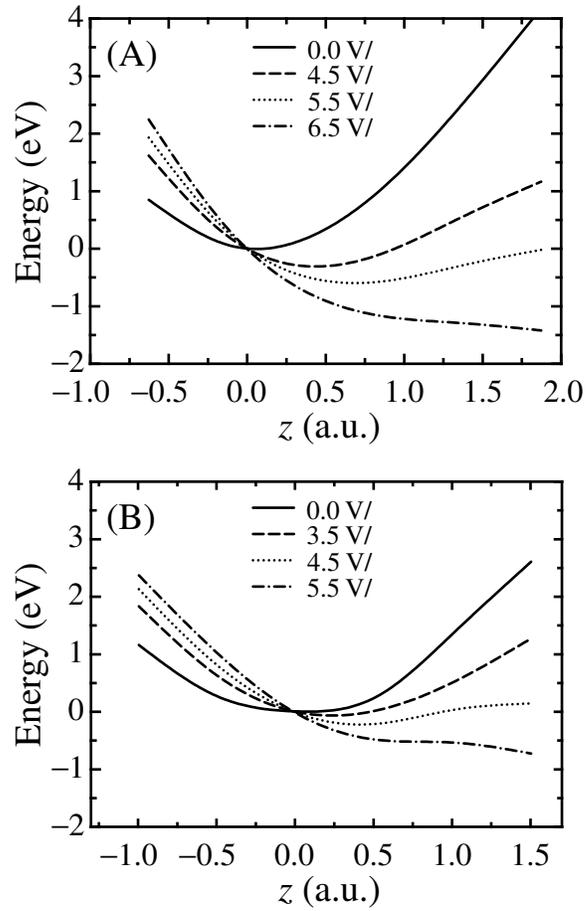}
\caption{Potential energies for evaporating atoms on (A) W(011) and (B) Mo(011) surfaces as a function of position $z$ of the evaporating atoms. The zero of the position $z$ and the energy are chosen to be those of the most stable atomic geometry in an absence of the external electric field.}
\label{fig:2}
\end{center}
\end{figure}

We now evaluate the activation energies for the field evaporation of surface atoms during the lifting of the surface atoms. One of the potentially important applications of the field-evaporation process is the direct determination of the binding strengths of surface atoms from the external electric field required for their removal. Figure \ref{fig:2} shows activation energies as a function of distance from the surface for various external electric fields. These energies become lower as the external electric fields are increased. On the basis of these results, the threshold values of the external electric fields for evaporation are $\sim$ 6 V/\AA \hspace{2mm} for tungsten and $\sim$ 5 V/\AA \hspace{2mm} for molybdenum, which agree with the experimental values of 5.7 V/\AA \hspace{2mm} for tungsten and 4.6 V/\AA \hspace{2mm} for molybdenum obtained using a field-ion microscope at low temperature. In addition, the threshold values of the external electric field are considered to vary depending on the bond strengths of the surface atoms, and the rate constants of field evaporation depend on the activation energies according to the Arrhenius formula \cite{Arrhenius}; the molybdenum surface atom is easily removed compared with the tungsten surface atom. This is consistent with the binding energy mentioned in the preceding section.

\subsection{Local-field enhancement}
\begin{figure}[htb]
\begin{center}
\includegraphics{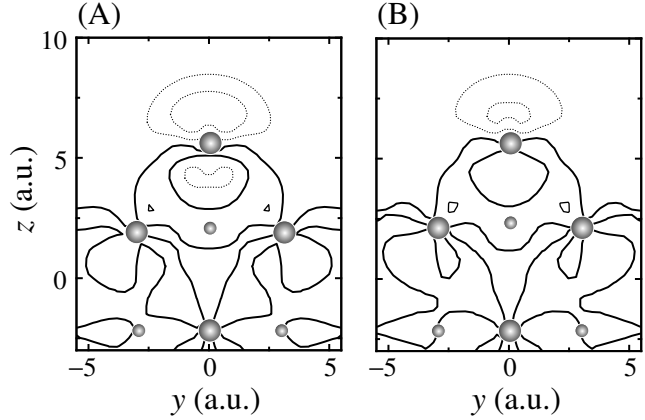}
\caption{Electronic density shifts $\rho ({\bf r},F)-\rho ({\bf r},F=0)$ on the (0$\bar{1}$1) cross sections containing the evaporating atoms on (A) tungsten and (B) molybdenum. External electric fields $F$ are 5.5 V/\AA \hspace{2mm} for tungsten and 4.5 V/\AA \hspace{2mm} for molybdenum. The contour spacing is 0.005 electron/supercell and thick curves represent 0.000 electron/supercell. Solid (dotted) curves represent positive (negative) values. Large and small balls indicate the atomic positions on and above the cross section, respectively.}
\label{fig:3}
\end{center}
\end{figure}

\begin{figure}[htb]
\begin{center}
\includegraphics{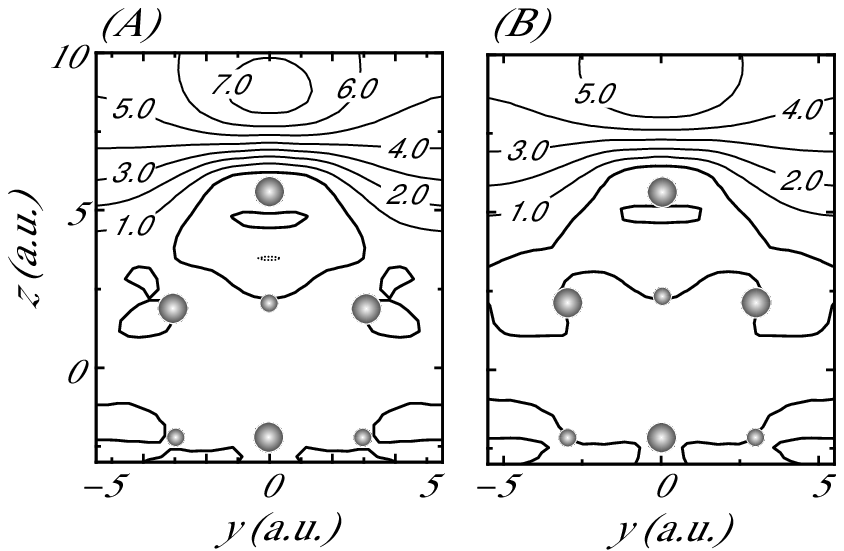}
\caption{Differences between the total electrostatic field in the presence of an external electric field and that in the absence of an electric field on the (010) cross sections containing the evaporating atoms on (A) tungsten and (B) molybdenum. External electric fields $F$ are 5.5 V/\AA \hspace{2mm} for tungsten and 4.5 V/\AA \hspace{2mm} for molybdenum. The contour spacing is 1.0 V/\AA \hspace{2mm} and and thick curves represent 0.0 V/\AA. Solid (dotted) curves represent positive (negative) values. Symbols have the same meanings as those in Fig. \ref{fig:3}.}
\label{fig:4}
\end{center}
\end{figure}

We depict in Fig. \ref{fig:3} the electron distributions $\rho ({\bf r},F)-\rho ({\bf r},F=0)$ induced by the external electric field \cite{comment3}. The overall charge around the surface atoms decreases, and the atoms are expected to be positive ions when they evaporate \cite{comment1}. We show in Fig. \ref{fig:4} the counter plots of the external electrostatic field as the differences between the total electrostatic field in the presence of an external electric field and that in the absence of an external electric field \cite{comment3}, $d(V({\bf r},F)-V({\bf r},F=0))/dz$, where $V$ is the sum of the external and Hartree potential functions. One can clearly recognize the expulsion of the external electronic field from the inside of the surface which is well known in the case of macroscopic metallic systems. However the LFE occurs not around the evaporating atoms but $\sim$ 3 a.u. above them, and there are no further significant differences in external electrostatic field around the evaporating atoms. In addition, the strength of the local field is much smaller than that expected on the basis of the classical model (16.5 V/\AA \hspace{2mm} for tungsten and 13.5 V/\AA \hspace{2mm} for molybdenum) \cite{rose}. This situation is consistent with the result obtained using a jellium surface \cite{lang1}.

\section{Conclusion}
\label{sec:concl}
We have studied the field-evaporation processes of surface atoms from atomically flat W(011) and Mo(001) surfaces. The threshold values of the external electric field for evaporation of the surface atoms, which are $\sim$ 6 V/\AA \hspace{2mm} for tungsten and $\sim$ 5 V/\AA \hspace{2mm} for molybdenum, are in agreement with the experimental results. They depend on the binding energy between the evaporating atoms and the surface. Moreover, as the external electric field increases, the activation energy for field evaporation becomes lower. While field evaporation has been believed to occur as a result of remarkable LFE around the evaporating atoms, we observe the LFE $\sim$ 3 a.u. above the atoms and the strength of the local field is not particularly high compared with those expected on the basis of the classical model.

\section*{Acknowledgements}
This research was supported by a Grant-in-Aid for the 21st Century COE ``Center for Atomistic Fabrication Technology'' and also by a Grant-in-Aid for Young Scientists (B) (Grant No. 14750022) from the Ministry of Education, Culture, Sports, Science and Technology. The numerical calculation was carried out with the computer facilities at the Institute for Solid State Physics at the University of Tokyo, and the Information Synergy Center at Tohoku University.




\begin{center}
\begin{table}
\caption{Binding energies as the atoms are displaced at sites A-C.}
\begin{tabular}{ccc|cc}
\hline \hline 
\multicolumn{2}{c}{W} && \multicolumn{2}{c}{Mo}  \\ \hline
Site & Binding energy (eV) && Site & Binding energy (eV) \\ \hline
A & 5.09 && A & 3.93 \\
B & 7.88 && B & 5.65 \\
C & 6.80 && C & 4.94 \\
\hline \hline
\end{tabular}
\label{tbl:1}
\end{table}
\end{center}

\newpage

\begin{thebibliography}{}


\bibitem{a}
See for example D.M. Eigler and E.K. Schweizer in Nature {\bf 344} (1990) 524, and I.-W. Lyo and P. Avouris in Science {\bf 253} (1991) 173.
\bibitem{tsong}
T.T. Tsong, {\it Atom Probe Field Ion Microscopy}, (Cambridge University Press, Cambridge, 1990), and references are therein.
\bibitem{Arrhenius}
S. Arrhenius, Z. Phys. Chem. {\bf 4} (1889) 226.
\bibitem{watanabe}
K. Watanabe and T. Satoh, Surf. Sci. {\bf 287-288} (1993) 502; T. Kawai, and K. Watanabe, Surf. Sci. {\bf 357-358} (1996) 830; T. Kawai, and K. Watanabe, Surf. Sci. {\bf 382} (1997) 320; T. Kawai, K. Watanabe, and K. Kobayashi, Ultramicroscopy {\bf 73} (1998) 205.
\bibitem{ono}
T. Ono and K. Hirose, J. Appl. Phys. {\bf 95} (2004) 1568.
\bibitem{lang1}
H.J. Kreuser, L.C. Wang, and N.D. Lang, Phys. Rev. B {\bf 45} (1992) 12050.
\bibitem{lang2}
N.D. Lang, Phys. Rev. B {\bf 45} (1992) 13599.
\bibitem{wang}
R.L.C. Wang, H.J. Kreuzer, R.G. Forbes, Surf. Sci. {\bf 350} (1996) 183.
\bibitem{chelikowsky1}
J.R. Chelikowsky, N. Troullier, and Y. Saad, Phys. Rev. Lett. {\bf 72} (1994) 1240.
\bibitem{chelikowsky2}
J.R. Chelikowsky, N. Troullier, K. Wu, and Y. Saad, Phys. Rev. B {\bf 50} (1994) 11355.
\bibitem{tsdg}
T. Ono and K. Hirose, Phys. Rev. Lett. {\bf 82} (1999) 5016.
\bibitem{hks2}
W. Kohn and L.J. Sham, Phys. Rev. {\bf 140} (1965) A1133.
\bibitem{pseudop}
N. Troullier and J.L. Martins, Phys. Rev. B {\bf 43} (1991) 1993.
\bibitem{kobayashi}
We used the norm-conserving pseudopotentials NCPS97 constructed by K. Kobayashi. See K. Kobayashi, Comput. Mater. Sci. {\bf 14} (1999) 72.
\bibitem{kleinman-bylander}
L. Kleinman and D.M. Bylander, Phys. Rev. Lett. {\bf 48} (1982) 1425.
\bibitem{lda}
J.P. Perdew and A. Zunger, Phys. Rev. B {\bf 23} (1981) 5048.
\bibitem{hks1}
P. Hohenberg and W. Kohn, Phys. Rev. {\bf 136} (1964) B864.
\bibitem{comment3}
In Figs. \ref{fig:3} and \ref{fig:4}, the charge distribution $\rho ({\bf r},F)$ and potential $V({\bf r},F)$ are generated using the stable atomic configuration at $F$=0.0 V/\AA.
\bibitem{comment1}
Due to the nature of the density functional theory \cite{hks1}, electrons necessarily transfer from the evaporating atom with a high potential to the surface with a low potential under an external electric field even though the distance between them becomes very large, e.g., $>5$ \AA. Therefore, the charge state of the evaporating atoms cannot be determined exactly by using the conventional formalism of the density functional theory.
\bibitem{rose}
D.J. Rose, J. Appl. Phys. {\bf 27} (1956) 215.
\end{thebibliography}
\end{document}